\documentclass[11pt]{article}
\pdfoutput=1

\usepackage[T1]{fontenc}
\usepackage[utf8]{inputenc}
\usepackage{microtype}
\usepackage{pdflscape}
\usepackage{multirow}
\usepackage{cellspace} 
\usepackage{xcolor,comment}
\definecolor{tabblue}{HTML}{1F77B4}  % tab:blue
\definecolor{taborange}{HTML}{FF7F0E} % tab:orange
\colorlet{tabblueDark}{tabblue!70!black}
\colorlet{taborangeDark}{taborange!70!black}
\usepackage{eso-pic}

\usepackage{amsmath,amssymb,mathtools}
\allowdisplaybreaks
\usepackage{tensor}
\usepackage{bbm}
\usepackage{braket}
\usepackage{rotating}

\usepackage{fullpage}
\usepackage[colorlinks=true,
            linkcolor=tabblueDark,
            citecolor=tabblueDark,
            urlcolor=tabblueDark]{hyperref}
\usepackage{cleveref} % optional but handy for refs
\usepackage{pdflscape}

\usepackage{graphicx}
\usepackage{float}
\usepackage{subcaption}

\graphicspath{{img/}}

\usepackage{cite}

\usepackage{authblk}

\allowdisplaybreaks

\renewcommand{\vec}[1]{\mbox{\boldmath$ #1 $}}

\newcommand{\beq}{\begin{equation}}
\newcommand{\eeq}{\end{equation}}

\renewcommand{\[}{\begin{equation}\begin{aligned}}
\renewcommand{\]}{\end{aligned}\end{equation}}

\usepackage{tcolorbox}
\definecolor{airforceblue}{rgb}{0.36, 0.54, 0.66}
\definecolor{azure}{rgb}{0.0, 0.5, 1.0}
\newtcolorbox{tdbox}{colback=airforceblue!40!white,colframe=azure!90!black} 

\allowdisplaybreaks
\numberwithin{equation}{section}

\title{\bf Non-local nonstabiliserness in Gluon and Graviton Scattering}

\author[1]{John Gargalionis\thanks{\href{mailto:john.gargalionis@adelaide.edu.au}{john.gargalionis@adelaide.edu.au}}}
\author[2]{Nathan Moynihan\thanks{\href{mailto:n.moynihan@qmul.ac.uk}{n.moynihan@qmul.ac.uk}}}
\author[2]{Michael L. Reichenberg Ashby\thanks{\href{mailto:m.l.reichenbergashby@qmul.ac.uk}{m.l.reichenbergashby@qmul.ac.uk}}}
\author[1]{Ewan N.\ V.\ Wallace\thanks{\href{mailto:ewan.n.wallace@adelaide.edu.au}{ewan.n.wallace@adelaide.edu.au}}}
\author[2]{Chris D.\ White\thanks{\href{mailto:christopher.white@qmul.ac.uk}{christopher.white@qmul.ac.uk}}}
\author[1]{Martin J.\ White\thanks{\href{mailto:martin.white@adelaide.edu.au}{martin.white@adelaide.edu.au}}}

\affil[1]{\it ARC Centre of Excellence for Dark Matter Particle Physics \& CSSM, Department of Physics, University of Adelaide, Adelaide, SA 5005, Australia}
\affil[2]{\it Centre for Theoretical Physics, School of Physical and Chemical Sciences, Queen Mary University of London, 327 Mile End Road, London E1 4NS, UK}

\date{{}}

\begin{document}

%\AddToShipoutPictureFG*{%
  % Absolute placement using picture coordinates (reliable)
  %\put(\LenToUnit{\paperwidth-1in},\LenToUnit{\paperheight-0.8in}){%
   % \makebox(0,0)[tr]{\preprintnum}%
  %}%
%}

\maketitle

\begin{abstract} 
The property of non-stabiliserness, or ``magic'', is of interest in quantum computing due to its role in developing fault-tolerant quantum algorithms with genuine computational advantage over classical counterparts. There has been much interest in quantifying magic in various physical systems, in order to probe how to produce and enhance it. The production of magic has previously been quantified in gluon and graviton scattering, in the so-called {\it helicity basis} relating particle spins with momentum directions. For a basis-independent statement, one should instead use the recently developed concept of {\it non-local non-stabiliserness}, and our aim in this paper is to derive how this varies for gluon and graviton scattering  processes. Our results show that, for many initial states, including those produced with polarised beams, the helicity basis coincides with a basis in which the non-local magic is manifest, providing a physical motivation for using the helicity basis to study quantum information quantities. However, this property breaks upon adding additional operators to the Yang-Mills Lagrangian, as would be the case in new physics scenarios.
\end{abstract}

\section{Introduction}
\label{sec:intro}
Bipartite quantum systems are characterised by distinctly quantum correlations that exist between the two subsystems, for which there is no classical counterpart. For example, two subsystems of a larger system are \emph{entangled} if the system cannot be decomposed into a product state of the subsystems. To give another example, certain states called \emph{stabiliser states} have been identified that, when acted on by quantum logic gates, confer no advantage over a classical computer. One can then define measures of \emph{non-stabiliserness} (or ``magic'') to rank those states that permit quantum advantage in computation. Various measures exist to quantify entanglement and non-stabiliserness, with the proposal of new measures and new methods for calculating existing measures being an active area of research.

The precise relationship between nonstabiliserness and entanglement remains an open problem in general. For a system composed of two qubits, the stabiliser states are either non-entangled or maximally-entangled states, indicating that the relationship is non-trivial. This is further illustrated by the fact that entanglement is invariant under local unitary transformations, whilst nonstabiliserness \emph{can} be altered by local transformations, and is thus a basis-dependent quantity. A recently proposed quantity that encapsulates knowledge of both entanglement and nonstabiliserness is the non-local nonstabiliserness which, for bipartite states, removes the basis dependence by performing a minimisation of any chosen magic measure over all possible local unitary transformations of the two subsystems~\cite{cao2025gravitationalbackreactionmagical,cao2024nontrivialareaoperatorsrequire,Qian:2025oit}. It is interesting to apply this measure to previous systems whose entanglement and magic properties have been well-studied, in order to gain further insights and tease out the deeper meaning of their observable quantum correlations. 

Much recent literature has examined the entanglement and nonstabiliserness properties of particle scattering and decay processes~\cite{Balasubramanian:2011wt, Seki:2014cgq, Peschanski:2016hgk,Grignani:2016igg,Kharzeev:2017qzs,Fan:2017hcd,Fan:2017mth,Cervera-Lierta:2017tdt,Beane:2018oxh,Rigobello:2021fxw,Low:2021ufv,Liu:2022grf,Fedida:2022izl,Cheung:2023hkq,Carena:2023vjc,Aoude:2024xpx,Low:2024mrk,Low:2024hvn,Thaler:2024anb,McGinnis:2025brt,Carena:2025wyh,Liu:2025pny,Hu:2025lua,Sou:2025tyf}. Of particular relevance here are studies of $2\to 2$ gluon scattering (described by Yang-Mills theory) and $2\to 2$ graviton scattering (described by general relativity), including a supersymmetric extension of both theories. In all cases, the massless nature of the particles being scattered ensures that they have two polarisation states, which in turn means that the initial and final states can be treated as a two-qubit system. One can then explore the ability of the scattering process to modify or generate quantum correlations in spin space. Ref.~\cite{Gargalionis:2025iqs} explored the nonstabiliserness of these systems in the helicity basis, comparing and contrasting this with the entanglement behaviour. For completeness, we note that other examples of the application of quantum information concepts to high energy physics examples can be found in Refs.~\cite{Afik:2020onf,Dong:2023xiw,Aoude:2022imd,Fabbrichesi:2021npl,Severi:2021cnj,Afik:2022kwm,Aguilar-Saavedra:2022uye,Fabbrichesi:2022ovb,Afik:2022dgh,Severi:2022qjy,Aguilar-Saavedra:2023hss,Han:2023fci,Simpson:2024hbr,Aguilar-Saavedra:2024hwd,Maltoni:2024csn,ATLAS:2023jzs,ATLAS:2023fsd,CMS:2024hgo,Barr:2024djo,Abel:1992kz,Abel:2025skj,Bechtle:2025ugc,Low:2025aqq,White:2024nuc,Liu:2025frx,Liu:2025qfl,Fabbrichesi:2025ywl,CMS:2025cim,Afik:2025ejh,Aoude:2025jzc,Barr:2022wyq,AshbyPickering:2022umy,Aguilar-Saavedra:2022wam,Aguilar-Saavedra:2022mpg,Fabbri:2023ncz,Aoude:2023hxv,Fabbrichesi:2023cev,Sakurai:2023nsc,Altomonte:2023mug,Afik:2024uif,Aguilar-Saavedra:2024vpd,Aguilar-Saavedra:2024whi,Grabarczyk:2024wnk,Morales:2024jhj,Altomonte:2024upf,Han:2024ugl,Cheng:2024rxi,Subba:2024aut,Subba:2024mnl,Horodecki:2025tpn,DelGratta:2025qyp,Nason:2025hix,Grossi:2024jae,Cheng:2025cuv,Ghodrati:2023uef}. Studies of magic in a variety of quantum systems can be found in Refs.~\cite{Beverland_2020,PhysRevApplied.19.034052,Leone:2023avk,Qassim2021improvedupperbounds,Leone:2021rzd,Haug:2023ffp,Magic1,PhysRevA.108.042408,Gu:2023qqq,Tirrito:2023fnw,Turkeshi:2023lqu,Leone:2024lfr,Tarabunga:2023ggd,Frau:2024qmf,Lami:2024osd,Robin:2024bdz,Chernyshev:2024pqy,White:2024nuc,Liu:2025frx,Liu:2025qfl,Fabbrichesi:2025ywl,Aoude:2025jzc,Busoni:2025dns,CMS:2025cim,Guo:2026yhz}.

In this paper, we will map the non-local nonstabiliserness, in order to tease out further insights on the observed pattern of quantum correlations. We believe this to be useful given the following motivation. It is known that the helicity basis, with its strong physical association of spin directions with particle directions in spacetime, can be optimal for spin correlation measurements and related observables. This suggests that there is a strong physical foundation for using the helicity basis to measure nonstabiliserness, such that one need not worry too much about the basis dependence of the magic in practice. In order to make this concrete, however, one should in fact check whether or not the helicity basis for the final state coincides with a basis in which the nonlocal magic is manifest. In this way, one can explicitly verify whether or not the helicity basis is optimal for probing non-stabiliserness.

In what follows, we will group initial stabiliser states into different categories, and see that for many categories, the magic in the helicity basis indeed coincides with the non-local magic, in agreement with the above expectation. For other types of initial state, however, the magic in the helicity basis is not in fact optimal, such that the minimum magic is obtained in an alternative choice of basis. As in ref.~\cite{Gargalionis:2025iqs}, we will consider results for massless particles of differing spin, where results for gluons (spin 1) can be straightforwardly recycled to obtain results for gluinos (spin 1/2), gravitinos (spin 3/2) and gravitons (spin 2). These secondary results follow from supersymmetric Ward identities and the KLT relations~\cite{Kawai:1985xq}, and the kinematic dependence inherent in these relations will allow us to explain differing profiles of non-local magic for different spins. Our findings may be portable to other quantum systems, particularly those involving massless excitations (e.g. near phase transitions). 

We will further probe the usefulness of the helicity basis by considering a variant of Yang-Mills (gluon) theory, in which one adds a higher-dimensional operator to the Lagrangian. Such operators generically arise in new physics corrections to the Standard Model of particle physics (see e.g. ref.~\cite{Aoude:2025jzc} for a recent discussion of this in the context of quantum non-stabiliserness), and adding such an operator here allows us to consider a one-parameter deformation of pure Yang-Mills theory. We will find in this case that the helicity basis is in fact no longer such that final states have the non-local magic manifest in general, even for simple initial states which are typically used in collider experiments with polarised beams. Thus, the usefulness of the helicity basis can be highly sensitive to the particular theory being talked about, underlining the importance of using non-local vs. local non-stabiliserness measures depending on the context. 

The structure of our paper is as follows. In section~\ref{sec:review}, we review relevant ideas regarding non-local magic~\cite{cao2025gravitationalbackreactionmagical,cao2024nontrivialareaoperatorsrequire,Qian:2025oit}, and the analysis of (basis-dependent) magic in scattering processes of Ref.~\cite{Gargalionis:2025iqs}. In section~\ref{sec:nonlocal}, we derive analytic results for the non-local magic for all possible initial stabiliser states in a $2\rightarrow2$ scattering process. We also generically show that, for many categories of initial state, the helicity basis is such that the non-local magic is manifest. In section~\ref{sec:results}, we present explicit results for the scattering of particles of various spins, and interpret a number of interesting features. In section~\ref{sec:EFT}, we examine the above-mentioned deformation of Yang-Mills theory. We discuss our results and conclude in section~\ref{sec:conclude}.

\section{Review of necessary background}
\label{sec:review}

In this section, we briefly review the concept of nonstabiliserness and its non-local generalisation, plus our approach to treating particle scattering as a two qubit system. The description of a system of two qubits starts with the assignment of basis states for each qubit $|0\rangle$ and $|1\rangle$, such that a basis for the full $2$-qubit
Hilbert space is provided by the states $|i\rangle\otimes|j\rangle\,$. A special set of states called \emph{stabiliser states} can be obtained by acting on the state $|0\rangle\otimes|0\rangle$ with particular gates that are within the \emph{Clifford group} $\mathcal{C}_2$, defined via:
\begin{equation}
    \mathcal{C}_2 \equiv \{U \in \text{U}(4):U P U^\dagger = e^{i\theta} P'\},
\end{equation}
Here, $\theta\in\{0,\pi/2,\pi,3\pi/2\}$ and $P$ is a \emph{Pauli string} operator defined in terms of Pauli operators or the identity matrix acting on each individual qubit,
\begin{equation}
  {\cal P}_n=P_A\otimes P_B,\quad
  P_X\in\{\mathbbm{1},\sigma_1,\sigma_2,\sigma_3\}\,.
  \label{Paulistring}
\end{equation}
such that a Pauli or $2\times 2$ identity matrix acts on each
individual qubit. The Pauli strings generate the \emph{Pauli group}, which consists of the Pauli strings weighted by phases of $\pm 1,\pm i$. 
Then, the {\it Pauli spectrum} of a general two-qubit quantum state $|\psi\rangle$ is defined as the set of expectation values of the Pauli strings
\begin{equation}
  {\rm spec}(|\psi\rangle)=\{\langle\psi|P|\psi\rangle,\quad P\in
  {\cal P}_n\}\,.
  \label{Paulispec}
\end{equation}
For stabiliser states, this spectrum has a particularly simple form, with 4 values equal to $\pm1$, and the rest zero. The relevance of stabiliser states for quantum computation results from the \emph{Gottesmann--Knill theorem}, which states that
a quantum circuit comprised of only Clifford circuits is no more efficient than a classical computer. It is therefore the ``nonstabiliserness'' (also known as ``magic'') of a quantum
state $|\psi\rangle$, which confers quantum advantage in computation, and there are a number of metrics that can be used to quantify how nonstabiliser a particular state is. Such measures should ideally be additive, whilst giving precisely zero for stabiliser states. 

In Ref.~\cite{Gargalionis:2025iqs}, we explored the {\it Stabiliser R\'{e}nyi Entropies} (SRE) of Ref.~\cite{Leone:2021rzd}, calculating how much of this quantity could be produced in the various gluon, graviton, gluino and gravitino scattering processes. For the pure states explored in this work, these can be written for $n$ qubits as
\begin{equation}
  M_q(|\psi\rangle)=\frac{1}{1-q}\log_2\left(\sum_{P\in{\cal P}_n}\frac{ \langle \psi|
  P|\psi\rangle^{2q} }{2^n}\right).
  \label{Mqdef}
\end{equation}
The integer $q\geq 2$ can be freely chosen, with each choice providing a different moment of the Pauli spectrum. In practice, one often chooses to use the Second Stabiliser R\'{e}nyi entropy (SSRE) as a summary statistic to quantify non-zero magic. A frequent criticism of the SSRE is that it is a basis-dependent quantity, since it is not invariant under local unitary transformations applied separately to each qubit. A possible way around this is to define the non-local magic by explicitly minimising the SSRE over independent basis transformations for the two qubits to give
\begin{equation}
\mathcal{M}_{AB}(|\psi\rangle)=\min_{U_A\otimes U_B} M_2(U_A\otimes U_B|\psi\rangle)
\end{equation}
We note that, although we have calculated the non-local magic using the SSRE, the definition can be used with any valid measure $M$ of nonstabiliserness~\cite{cao2025gravitationalbackreactionmagical,cao2024nontrivialareaoperatorsrequire,Qian:2025oit}. For the special case of 2-qubit pure states, the non-local magic can be computed analytically. The starting point is a result derived in Ref.~\cite{Qian:2025oit}, namely that the non-local magic of a 2-qubit state of the form 
\begin{equation}
|\psi\rangle=\cos(\theta)|00\rangle+\sin(\theta)|11\rangle.
\label{Schmidt}
\end{equation}
is
\begin{equation}
\mathcal{M}_{AB}(|\psi\rangle)=\log_2\left(\frac{8}{7+\cos(8\theta)} \right).
\label{MABexact}
\end{equation}
Equation~(\ref{Schmidt}) is known as a {\it Schmidt decomposition}, and is possible for any 2-qubit state. To see this, note that a general 2-qubit state can be written as
    \begin{equation}
        \ket\psi = \sum_{iI} \psi^{iI} \ket{i}_A\otimes\ket{I}_B
        \label{psiform}
    \end{equation}
in the computational basis, where $\{\psi^{iI}\}$ are components of a complex $2\times 2$ matrix. The latter has a singular value decomposition
    \begin{equation}
        \psi^{iI}={U^i}_j \Sigma^{jJ} {(V^\dagger)_J}^I,
        \label{SVD}
    \end{equation}
where the two matrices ${U^i}_j$ and ${(V^\dagger)_J}^I$ can then be seen as local unitary transformations acting on each individual qubit. The diagonal matrix $\Sigma^{jJ}={\rm diag}(\sqrt{\eta_1},\sqrt{\eta_2})$ is in terms of the eigenvalues of the reduced density matrix  
    \begin{equation}
    \rho_A = \text{tr}_B \ket\psi\bra\psi=\sum_{ij}\psi^{i I}(\psi^\dagger)_{I j}\ket{i}_A\otimes\bra{j}_A,
    \label{rhoA}
    \end{equation}
whose eigenspectrum is known as the {\it entanglement spectrum} of the quantum state. On general grounds, the entanglement spectrum can always be written as $(\cos^2\theta,\sin^2\theta)$, such that the diagonally transformed state has the form
    \begin{equation}
        \ket\Sigma=\cos\theta \ket{00}+\sin\theta \ket{11}=\sqrt{\eta_1}\ket{00}+\sqrt{\eta_2}\ket{11},
        \label{Sigmaform}
    \end{equation}
which is precisely the Schmidt decomposition of eq.~(\ref{Schmidt}), where we have defined
\begin{equation}
(\eta_1,\eta_2)=(\cos^2\theta,\sin^2\theta).
\label{eta12def}
\end{equation}
To see that such a basis locally manifests the non-local magic, one computes the matrix
\begin{equation}
    R_{\mu\nu} = \bra{\Sigma}\sigma_\mu\otimes\sigma_\nu \ket{\Sigma} = \left( \begin{matrix} 1& \vec v_B^T\\\vec v_A & T_{AB} \end{matrix} \right) =\left(\begin{matrix}1&0&0&\cos 2\theta\\0&\sin2\theta&0&0\\0&0&-\sin 2\theta&0\\\cos 2\theta&0&0&1\end{matrix}\right),
\end{equation}
where $\vec v_A$ and $ \vec v_B$ are the Bloch vectors corresponding to the reduced density matrices of the single-qubit subsystems and $T_{AB}$ represents the correlation matrix.
Under a local unitary transformation $U_A \otimes U_B$, $R_{\mu\nu}$ transforms in the adjoint by orthogonal matrices $O_A,O_B$:
\begin{equation}
    \vec v_A\rightarrow O_A \vec v_A,\,\vec v_B \rightarrow O_B\vec v_B,\, T_{AB}\rightarrow O_A T_{AB}O_B^T.
\end{equation}
Given that the action of any orthogonal matrix will preserve the $\ell^2$-norm of the Bloch vectors, and $\sum_i|\vec v^i|^4\le\sum_i|\vec v^i|^2$, the contribution to the magic from the $R_{i0}$ and $R_{0j}$ components is clearly minimised when the Bloch vectors are fully polarised. Similarly, the contribution from $R_{ij}$ is minimised when $T_{AB}$ is diagonal. Therefore, a pure 2-qubit state expressed in its Schmidt basis will locally realise its non-local magic. This in turn gives an interpretation of the non-local magic directly in terms of the entanglement spectrum: from eqs.~(\ref{MABexact}, \ref{Sigmaform}), one finds (upon using appropriate trigonometric identities)
    \begin{equation}
        \mathcal{M}_{AB}(|\psi\rangle)=-\log_2 \left(1+4\eta_1(\eta_1-1)(1-2\eta_1)^2\right),
        \label{MABexact2}
    \end{equation}
where $\eta_1=\cos^2\theta$. Alternatively, the entanglement spectrum  for a 2-qubit state can be expressed entirely in terms of the basis-independent {\it concurrence}, an entanglement measure defined by 
\begin{equation}
    \chi=\sqrt{2\left(1-{\rm Tr}[(\rho_A)^2]\right)}
    =\sqrt{2\left(1-{\rm Tr}[(\rho_B)^2]\right)},
    \label{chidef}
\end{equation}
where $\rho_{A,B}$ are the reduced density matrices obtained by tracing over qubits $B$ and $A$ respectively. One then finds:
\begin{equation}
    \mathcal M_{AB}(\ket \psi ) = -\log_2 (1-\chi^2+\chi^4),
    \label{MABchi}
\end{equation}
which will be useful later when we discuss the relationship between non-local magic and entanglement.

The above discussion provides a clear recipe for analytically constructing the non-local magic for a given 2-qubit state: from the state itself, one may obtain the reduced density matrix. Then, either eigenvalue of the latter can be used to directly obtain the non-local magic via eq.~(\ref{MABexact2}). In this paper, as in our previous Ref.~\cite{Gargalionis:2025iqs}, the 2-qubit states we will consider will be final states produced in $2\rightarrow2$ scattering processes of massless particles with non-zero spin. As described in detail there, the incoming state of such a process consists of states 
\begin{equation}
\ket{\text{in}}\sim\ket{J}\otimes\ket{a_1 a_2}\otimes\ket{\mathbf{p}_1\mathbf{p}_2},
\end{equation}
labelled by momenta $\{\mathbf{p}_i\}$ and helicities $J=(\lambda_1,\lambda_2)$, where $\lambda_i=\pm$ represents the component of the particle spin along the momentum direction. We also allow for possible charge (colour) degrees of freedom $\{a_i\}$. The final state is found acting with the scattering matrix ($S$-matrix) on a given initial state. We further assume that our measurement projects onto a state of definite momentum and colour charge (if relevant), so that the effect of the scattering is to produce a state
\begin{equation}
    \ket{\text{out}}\propto \sum_{K}i\mathcal{A}(J\to K)\ket{K;\cdots;\mathbf{p}_3\mathbf{p}_4},
    \label{outstate}
\end{equation}
where ${\cal A}$ is the scattering amplitude for definite initial and final helicity configurations $J$ and $K$. From now on, we will suppress explicit momentum and colour degrees of freedom, and note that the final state must be normalised according to
\begin{equation}
\ket{\psi}\rightarrow \frac{1}{\sqrt{\braket{\psi|\psi}}}\ket{\psi}.
\label{psinorm}
\end{equation}
Explicit results for the (leading order) helicity amplitudes for massless particles of spin 1/2 (gluinos), 1 (gluons), 3/2 (gravitinos) and 2 (gravitons) are well-known, where the amplitudes for different spins can be obtained from the gluon cases using known relations from the scattering amplitude literature. For example, gluino and gluon amplitudes are related by supersymmetric Ward identities~\cite{Grisaru:1977px,Bianchi:2008pu,Dixon:2010ik,Grisaru:1976vm}. Once these are known, gravitino and graviton amplitudes can be obtained using the so-called {\it KLT relations}, originating from string theory~\cite{Kawai:1985xq}, and whose field theory analogue is known as the {\it double copy}~\cite{Bern:2008qj,Bern:2010yg} (see e.g. refs.~\cite{Bern:2019prr,White:2024pve} for pedagogical reviews). As stressed in Ref.~\cite{Gargalionis:2025iqs}, the convenience of studying these particular particles is that it comprises a family of 2-qubit systems, related by varying the spin. Importantly, the requirement for the particles to be massless ensures that there are only two polarisation states in each case. The same considerations also apply to the study of non-local magic, which we commence in the following section.

\section{Non-local magic in $2\rightarrow 2$ scattering}
\label{sec:nonlocal}

In the previous section, we saw that non-local magic can be calculated from a given quantum density matrix, directly in terms of the entanglement spectrum. We now spell out how this works explicitly for states of the form of eqs.~(\ref{outstate}, \ref{psinorm}), obtained from $2\rightarrow 2$ scattering. Furthermore, by leaving the amplitudes unspecified, our arguments in this section will apply to any of the particles mentioned above. Our starting point is to note the well-known property that, at leading order in perturbation theory, only certain helicity amplitudes are non-vanishing. If all momenta are outgoing, configurations in which all helicities are equal, or all equal but one, have vanishing amplitudes. The remaining amplitudes are referred to as {\it maximally helicity violating (MHV)}, and for $2\rightarrow 2$ scattering must therefore have two positive helicities, and two negative. Returning to the physical case of incoming initial momenta, and using the fact that momentum reversal flips the helicity, the structure of the amplitude matrix entering eq.~(\ref{outstate}) (i.e. in the computational basis $(\ket{++},\ket{+-},\ket{-+},\ket{--})$) is block diagonal:
\begin{equation}
    {\cal A}=\left(\begin{array}{cccc}
    {\cal A}(++\rightarrow ++) & 0 & 0 & 0\\
    0 & {\cal A}(+-\rightarrow +-) & {\cal A}(+-\rightarrow -+) & 0\\
    0 & {\cal A}(-+\rightarrow +-) & {\cal A}(-+\rightarrow -+) & 0\\
    0 & 0 & 0 & {\cal A}(--\rightarrow --)
    \end{array}\right).
    \label{ampmatrix}
\end{equation}
One also has the relations
\begin{align}
    \Gamma&\equiv {\cal A}(++\rightarrow ++)={\cal A}(--\rightarrow --),
    \notag\\
    \Sigma&\equiv {\cal A}(+-\rightarrow +-)={\cal A}(-+\rightarrow -+),
    \notag\\
    \Delta&\equiv {\cal A}(+-\rightarrow -+)={\cal A}(-+\rightarrow +-),
    \label{crossing}
\end{align}
so that we may further simplify eq.~(\ref{ampmatrix}) to
\begin{equation}
    {\cal A}=\left(\begin{array}{cccc}
    \Gamma & 0 & 0 & 0\\
    0 & \Sigma & \Delta & 0\\
    0 & \Delta & \Sigma & 0\\
    0 & 0 & 0 & \Gamma
    \end{array}\right).
    \label{ampmatrix2}
\end{equation}
In examining the production of magic, it is customary to take the initial state to be a given stabiliser state -- whose local or non-local magic is zero by definition -- and then quantify how much magic arises in the final state. To find the resulting density matrix for the final state, it is convenient to express the amplitude matrix of eq.~(\ref{ampmatrix2}) in terms of Pauli string operators, via the {\it Fano decomposition}
\begin{equation}
    {\cal A}=\frac12\Big[
    (\Gamma+\Sigma)\mathbbm{1}\otimes \mathbbm{1}+(\Gamma-\Sigma)\sigma_3\otimes \sigma_3
    +\Delta(\sigma_1\otimes \sigma_1+\sigma_2\otimes \sigma_2)
    \Big].
    \label{Fano}
\end{equation}
Given an initial state with density matrix $\rho_{\rm in}$, the density matrix for the corresponding out state is given by 
\begin{equation}
\rho_{\rm out}=\frac{{\cal A}\rho_{\rm in}{\cal A}^{\dagger}}
{{\rm Tr}[{\cal A}\rho_{\rm in}{\cal A}^{\dagger}]}.
\label{rhoout}
\end{equation}
The density matrix for an initial stabiliser state is also conveniently written in terms of Pauli strings, given the role the latter play in defining a stabiliser state in the first place. For two qubits, there are 60 possible stabiliser states, and a full characterisation of their density matrices is given in table~\ref{tab:stabstates}. In the table, we have grouped the states into distinct groups (labelled by roman numerals), corresponding to distinct values of the local magic in the helicity basis. In each group, there are two types of states. First, there are product states whose density matrix decomposes into the form
\begin{displaymath}
    \rho=\rho_A\otimes \rho_B.
\end{displaymath}
These are labelled by $\otimes$ in the table, and we use the symbol $\star$ to denote states which do not have a product form, and which are therefore entangled. From eqs.~(\ref{Fano}, \ref{rhoout}) and table~\ref{tab:stabstates}, one can find the Fano decomposition of the final state corresponding to any given initial stabiliser state. As an example, let us take one of the tensor product stabiliser states from group I($\otimes$), with density matrix
\begin{equation}
    \rho^{(1)}_{\rm in}=\frac{1}{4}(\mathbbm{1}+
         \sigma_1)\otimes(\mathbbm{1}+ \sigma_1).
         \label{rhoin1}
\end{equation}
By an explicit calculation according to the above procedure, we find the density matrix for the corresponding out state to be 
\begin{align}
    \rho^{(1)}_{\rm out}&=\frac14\left(\mathbbm{1}\otimes\mathbbm{1}+
    \sigma_1\otimes \sigma_1\right)+
    \frac{2\Gamma(\Delta+\Sigma)}{4(\Gamma^2+(\Delta+\Sigma)^2}
        \left(\sigma_1\otimes\mathbbm{1}+\mathbbm{1}\otimes\sigma_1 \right)\notag\\
        &\quad+\frac{\Gamma^2-(\Delta+\Sigma)^2}{4(\Gamma^2+(\Delta+\Sigma)^2}\left(\sigma_3\otimes\sigma_3
        -\sigma_2\otimes\sigma_2\right).
         \label{rhoout1}
\end{align}
The advantage of this representation is that it is now straightforward to read off the Pauli spectrum, which corresponds to the coefficients of each Pauli string operator. Substituting this into the SSRE from eq.~(\ref{Mqdef}) ($q=2$), we may write the magic as
\begin{equation}
    M_2=-\log_2\left(\zeta_2^{\rm local}\right),
    \label{M2def}
\end{equation}
where the argument for the particular state of eq.~(\ref{rhoout1}) is found to be
\begin{equation}
    \zeta_2^{\rm local}\left(\rho_\text{out}^{(1)}\right)=\frac{\Gamma^8+14\Gamma^4(\Sigma+\Delta)^4+(\Sigma+\Delta)^8}{(\Gamma^2+(\Sigma+\Delta)^2)^4}.
    \label{zeta2local1}
\end{equation}
As described in section~\ref{sec:review}, we can calculate the non-local magic using the eigenvalues of the reduced density matrix for the out-state. Writing this as
\begin{equation}
    {\cal M}_{AB}=-\log_2\left(\zeta_2^{\rm non-local}\right),
    \label{MABdef2}
\end{equation}
the argument for the state of eq.~(\ref{rhoout1}) turns out to be 
\begin{equation}
    \zeta_2^{\rm non-local}\left(\rho_\text{out}^{(1)}\right)=\frac{\Gamma^8+14\Gamma^4(\Sigma+\Delta)^4+(\Sigma+\Delta)^8
}{(\Gamma^2+(\Sigma+\Delta)^2)^4}.
    \label{zeta2nonlocal1}
\end{equation}
We thus see that, for the particular state considered here, the local and non-local magic are identical. In other words, the helicity basis provides a basis in which the non-local is manifest, without having to consider further local unitary transformations of each qubit. The question then arises of whether this remains true for the 59 other stabiliser states, and we report the complete set of results in table~\ref{tab:stabstates}, alongside each (sub-)group of stabiliser states. There are a number of noteworthy features in the table. Firstly we see that, whilst there is a common value of the local magic for each group, the same is not true for the non-local magic. Rather, there can be different values of the latter, depending on whether one considers a product ($\otimes$) or entangled ($\star$) state. Notably, all (maximally) entangled initial stabiliser states lead to vanishing non-local magic in the final state. Secondly, we may note that for several sub-groups, comprising 18 out of the total 60 stabiliser states, the local and non-local magic are the same. In other words, for many stabiliser states, the helicity basis is such that the non-local magic is manifest, requiring no additional minimisation over alternative bases. Furthermore, initial states that would actually arise in a collider with polarised beams (i.e. the product states $\ket{\pm \pm}$ and $\ket{\pm \mp}$) are such that their corresponding final states have manifest non-local magic in the helicity basis. This lends credence to the view that the helicity basis is a particularly well-motivated choice for the study of magic, a conclusion that may be useful when studying magic in more complex scattering processes, for which analytic results for non-local magic are unavailable, and where numerical minimisation over local unitary transformations proves computationally infeasible. It remains true, however, that for the majority of initial stabiliser states, the final state local magic in the helicity basis is not the same as the non-local magic.

\begin{sidewaystable}
        \centering
        \begin{tabular}{cScScScSlScSc}
           \multicolumn{2}{c}{Group} & Mult. & \multicolumn{2}{c}{$4\rho_\text{in}^{\text{stab.}}$} & $\zeta_2^\text{local}$ & $\zeta_2^\text{non-local}$ \\
           \hline\hline
            \multirow{3}{*}{I} & $\otimes$ & 4 & $(\mathbbm{1} \pm \sigma_i)\otimes (\mathbbm{1} \pm \sigma_i)$ & $i\in\{1,2\}$ & \multirow{3}{*}{$\frac{\Gamma^8+14\Gamma^4(\Sigma+\Delta)^4+(\Sigma+\Delta)^8}{(\Gamma^2+(\Sigma+\Delta)^2)^4}$} & $\frac{\Gamma^8+14\Gamma^4(\Sigma+\Delta)^4+(\Sigma+\Delta)^8}{\left(\Gamma^2+(\Sigma+\Delta)^2\right)^4}$\\
            & $\star$ & 4 & $\mathbbm 1^{\otimes 2}+{\sigma_i}^{\otimes 2} \pm(\sigma_j \otimes \sigma_3+\sigma_3\otimes \sigma_j)$ & $i,j\in\{1,2\}$ & & 1\\
            \hline
            \multirow{3}{*}{II} & $\otimes$ & 2 & $(\mathbbm1\pm \sigma_3)\otimes (\mathbbm 1\pm \sigma_3)$ & & \multirow{3}{*}{1}& \multirow{3}{*}{1}\\
            & $\star$ & 2 & $\mathbbm1^{\otimes 2} + {\sigma_3 }^{\otimes 2} \pm(\sigma_1\otimes\sigma_2+\sigma_2\otimes \sigma_1)$ & & &\\
            \hline
            \multirow{3}{*}{III} & $\otimes$ & 4 & $(\mathbbm1 \pm \sigma_i)\otimes (\mathbbm1 \mp \sigma_i)$ & $i\in\{1,2\}$&\multirow{3}{*}{$\frac{\Gamma^8+14\Gamma^4(\Sigma-\Delta)^4+(\Sigma-\Delta)^8}{(\Gamma^2+(\Sigma-\Delta)^2)^4}$} & $\frac{\Gamma^8+14\Gamma^4(\Sigma-\Delta)^4+(\Sigma-\Delta)^8}{(\Gamma^2+(\Sigma-\Delta)^2)^4}$\\
            & $\star$ & 4 & $\mathbbm 1^{\otimes 2}-{\sigma_i}^{\otimes 2} \pm(\sigma_j \otimes \sigma_3-\sigma_3\otimes \sigma_j)$ & $i,j\in\{1,2\}$& & 1\\
            \hline
            \multirow{3}{*}{IV} & $\otimes$ & 2 & $(\mathbbm1 \pm \sigma_3)\otimes(\mathbbm1\mp \sigma_3)$ & &\multirow{3}{*}{$\frac{\Sigma^8+14\Sigma^4\Delta^4+\Delta^8}{(\Sigma^2+\Delta^2)^4}$} & $\frac{\Sigma^8+14\Sigma^4\Delta^4+\Delta^8}{(\Sigma^2+\Delta^2)^4}$\\
            & $\star$ & 2 & $\mathbbm 1^{\otimes 2} -{\sigma_3}^{\otimes2} \pm (\sigma_1 \otimes\sigma_2 - \sigma_2 \otimes\sigma_1)$ & & & 1\\
            \hline
            \multirow{3}{*}{V} & $\otimes$ & 8 & $(\mathbbm1 + \iota \sigma_i)\otimes (\mathbbm1 + \kappa \sigma_j)$ & $i,j\in\{1,2\}$& \multirow{5}{*}{$\frac{\Gamma^8 +\Sigma^8 + \Delta^8+ 14 (\Gamma^4\Sigma^4+\Sigma^4\Delta^4 + \Delta^4\Gamma^4)}{(\Gamma^2+\Sigma^2+\Delta^2)^4}$}& $\frac{\Gamma^8+14\Gamma^4(\Sigma^2+\Delta^2)^2+(\Sigma^2+\Delta^2)^4}{(\Gamma^2+\Sigma^2+\Delta^2)^4}$\\
            & $\star$ & 8 & $\mathbbm 1^{\otimes 2}+ \iota \sigma_i\otimes\sigma_j + \kappa \sigma_j\otimes\sigma_k+\lambda \sigma_k\otimes\sigma_i$ & $\iota\kappa\lambda= -1$& &1\\
            \cline{1-5}\cline{7-0}
            VI & $\otimes$ & 16 & $(\mathbbm1 + \iota \sigma_i)\otimes(\mathbbm1 + \kappa \sigma_j)$ & $i=3 \,\text{OR}\, j=3$ & & $1+\frac{(2\Sigma\Delta)^4-(\Gamma^2+\Sigma^2+\Delta^2)^2(2\Sigma\Delta)^2}{(\Gamma^2+\Sigma^2+\Delta^2)^4}$\\
            \hline
            VII & \text{$\star$} & 4 & $\mathbbm1^{\otimes 2} + \iota {\sigma_1 }^{\otimes 2} + \kappa {\sigma_2 }^{\otimes 2}+\lambda {\sigma_3 }^{\otimes 2}$ & $\iota\kappa\lambda = -1$ & 1 & 1\\
        \end{tabular}
        \caption{Arguments of the local and non-local magic as defined in eqs.~(\ref{M2def}, \ref{MABdef2}) for the out states corresponding to the 60 pure stabiliser in states. The stabiliser states have been partitioned into seven groups by their structure and magic profiles, where within each group we further distinguish between product ($\otimes$) and entangled ($\star$) states. In describing each state, we have used the shorthand notation $A^{\otimes 2}\equiv A\otimes A$, and also introduced coefficients $\iota,\kappa,\lambda = \pm1$. The non-local magic of the outstates produced by the tensor product stabiliser states $(\otimes)$ from Groups I-IV is locally manifest in the helicity basis due to their only mixing $\mathbbm 1$ with one of the Pauli operators $\sigma_i$.}
        \label{tab:stabstates}
\end{sidewaystable}

A third feature of table~\ref{tab:stabstates} is that for two groups of stabiliser states, namely II and VII, the (non-)local magic vanishes (i.e. the argument of the logarithm in both cases is 1). The explanation for this is simply that the action of the amplitude matrix is such as to produce another stabiliser state. More interesting results are obtained for the other groups, where the amount of (non-)local magic in each case depends on the specific amplitudes involved, and thus upon the particle species. 

\section{Results}
\label{sec:results}

\begin{figure}
    \centering
    \scalebox{0.5}{\includegraphics{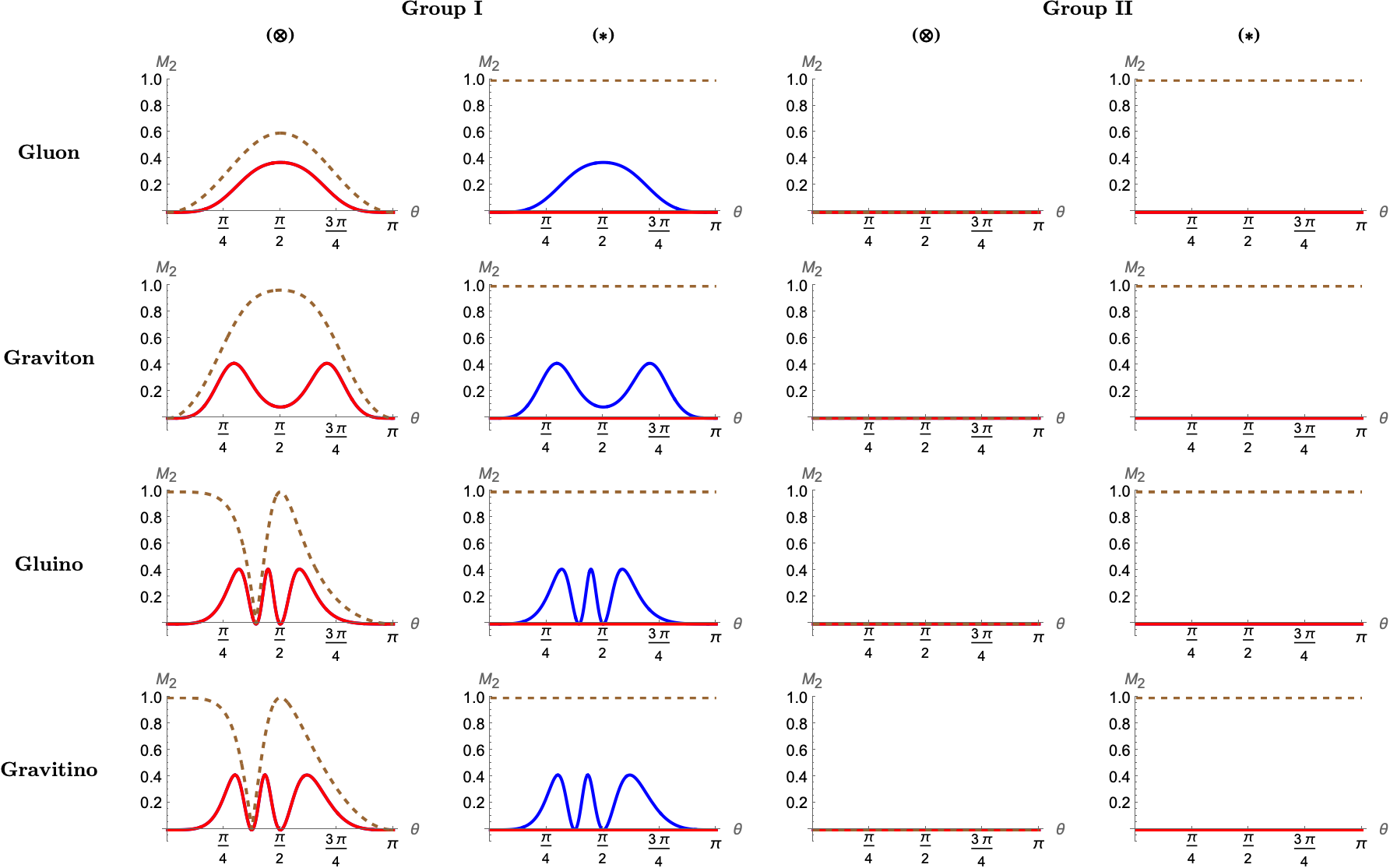}}
    \vspace{2em}
    \scalebox{0.5}{\includegraphics{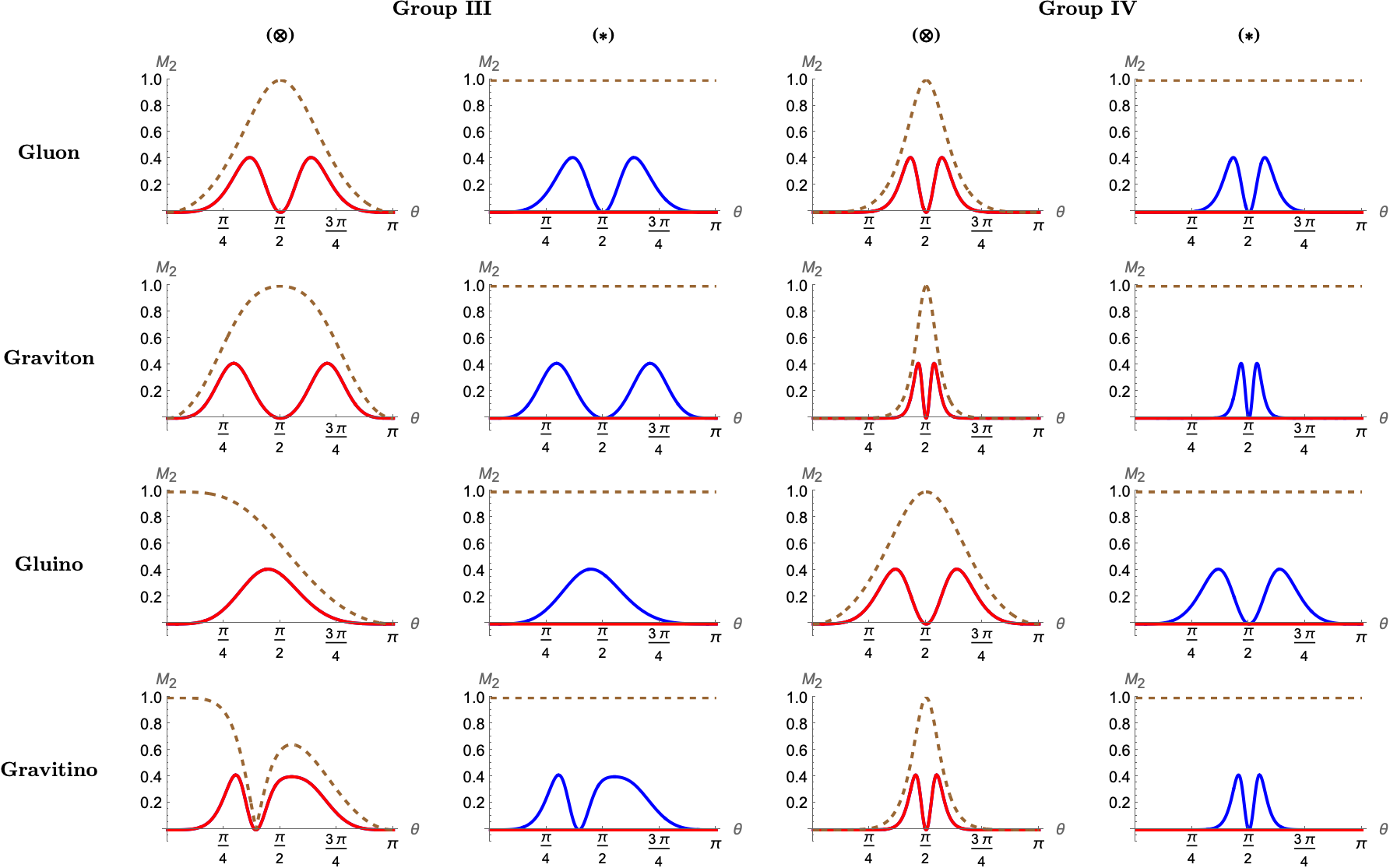}}
    \caption{The non-local magic (red) arising from representative initial stabiliser states from the groups in table~\ref{tab:stabstates}. Also shown is the local magic in the helicity basis (blue), where this differs from the non-local magic; and  the concurrence (dashed).}
    \label{fig:magic_plots_1}
\end{figure}

\begin{figure}
    \centering
    \scalebox{0.5}{\includegraphics{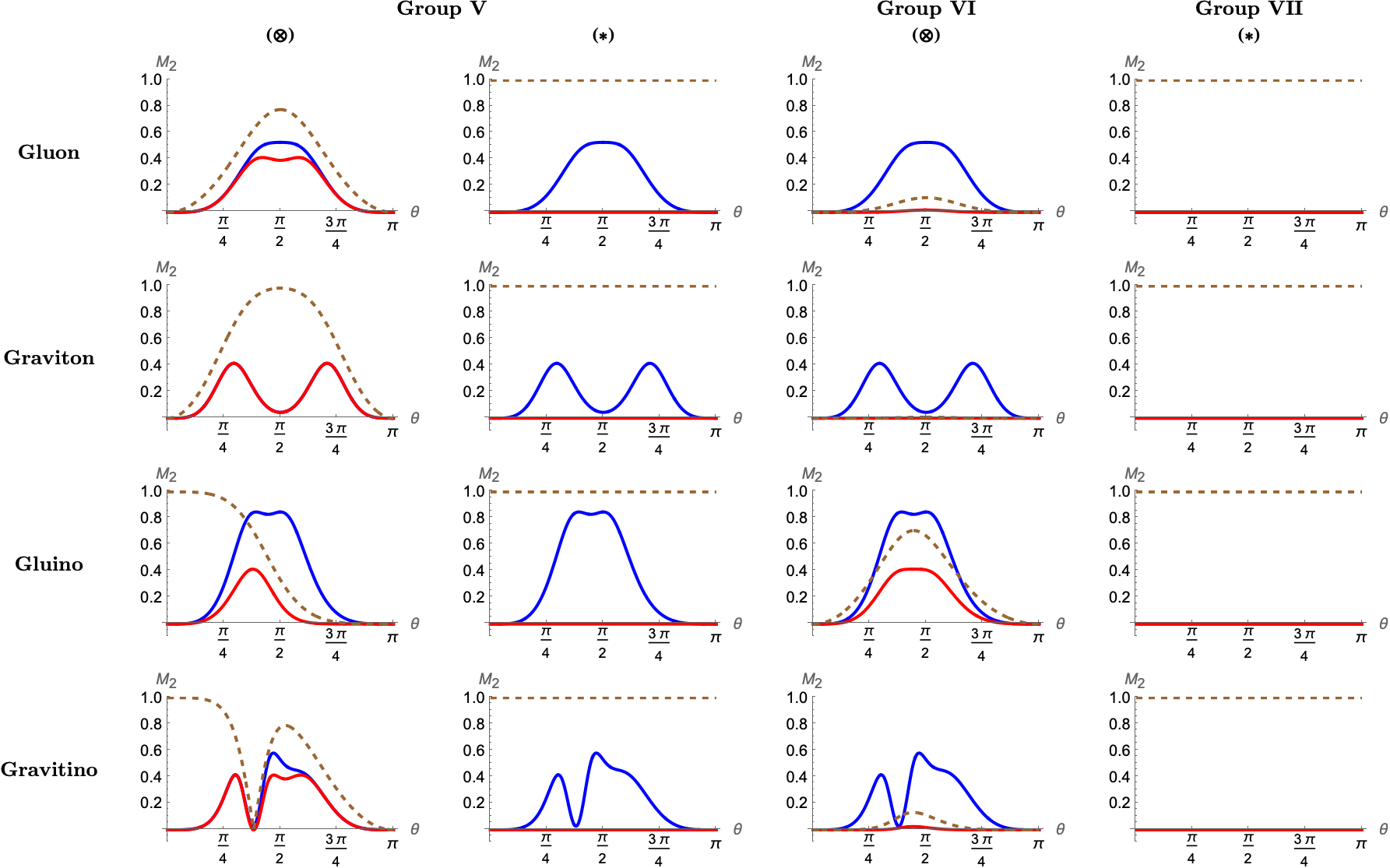}}
    \caption{The non-local magic (red) arising from representative initial stabiliser states from the groups in table~\ref{tab:stabstates}. Also shown is the local magic in the helicity basis (blue), where this differs from the non-local magic; and  the concurrence (dashed).}
    \label{fig:magic_plots_2}
\end{figure}

In the previous section, we have derived analytic results for the non-local magic produced from initial stabiliser states in $2\rightarrow2$ scattering of massless particles. Our results are written in terms of the general structure of the amplitudes matrix, independent of the spin. In this section, we collect results for massless particles of different spin, as considered for the local magic in Ref.~\cite{Gargalionis:2025iqs}. For convenience, we collect the relevant combinations of helicity amplitudes, in our present notation, in appendix~\ref{app:amplitudes}. Each amplitude is written in terms of the Mandelstam invariants
\begin{equation}
  s=(p_1+p_2)^2\,,\quad t=(p_1-p_3)^2\,,\quad u=(p_1-p_4)^2\,.
  \label{mandies}
\end{equation}
satisfying 
\begin{equation}
  s+t+u=0,
  \label{momcon}
\end{equation}
where $(p_1,p_2)$ and $(p_3,p_4)$ are the incoming and outgoing 4-momenta respectively. We furthermore work in the centre-of-mass frame, such that 
\begin{align}
  p_1&=(E,0,0,E)\,;\notag\\
  p_2&=(E,0,0,-E)\,;\notag\\
  p_3&=(E,E\sin\theta,0,E\cos\theta)\,;\notag\\
  p_4&=(E,-E\sin\theta,0,-E\cos\theta)\,,
  \label{pparam}
\end{align}
where $E$ is each particle's energy, and $\theta$ the
scattering angle. Substituting eq.~(\ref{pparam}) into eq.~(\ref{mandies}) yields
\begin{equation}
  s=4E^2\,,\quad t=-4E^2\sin^2\left(\frac{\theta}{2}\right)\,,\quad
  u=-4E^2\cos^2\left(\frac{\theta}{2}\right)\,,
  \label{mandies2}
\end{equation}
and the dimensionless nature of the argument of the (non-)local magic means that all energy dependence cancels out. Hence, the only kinematic dependence of the magic is through the scattering angle. This can clearly be seen in figs.~\ref{fig:magic_plots_1} and~\ref{fig:magic_plots_2}, which show the non-local magic (in red) and local magic in the helicity basis (in blue). Where the non-local and local magic are the same, the red curves lie entirely on top of the blue curves. Results are plotted using representative states from each of the groups in table~\ref{tab:stabstates}, and for particles of spin 1/2 (gluinos), spin 1 (gluons), spin 3/2 (gravitinos) and spin 2 (gravitons). The figures clearly illustrate the following behaviours noted in the previous section: (i) the (non-)local magic vanishes for groups I($\star$), II, III($\star$), IV($\star$), V($\star$) and VII; (ii) The (helicity basis) local, and non-local, magic are the same for groups I($\otimes$), II, III($\otimes$), IV($\otimes$), and VII. Furthermore, we see that for all groups where the non-local magic is non-zero, there is a non-trivial dependence on the scattering angle. For a given group of stabiliser states, this profile can gain or lose nodes as the spin increases, where the underlying mechanism for this is the fact that amplitudes of different spin are related by supersymmetric Ward identities and / or KLT relations. These relations involve multiplying amplitudes by ratios of Mandelstam invariants, whose angular dependence (from eq.~(\ref{mandies2})) then modifies the angular distributions of non-local magic. The particularly interesting cases are those for which the local and non-local magic are different. For group V($\otimes$), the most pronounced difference is for gluinos, with relatively minor differences for gluons, gravitons and gravitinos. There is a much larger effect for group VI($\otimes$), where one sees almost vanishing non-local magic for all species apart from gluinos, with a sizable amount of local magic in the helicity basis. 

In many cases, the (non-)local magic vanishes at intermediate values of the scattering angle, due to the presence of a stabiliser state there. One may then explore a recurring theme in the quantum information literature, namely the relationship between magic and entanglement. It is known that maximally entangled states are typically stabiliser states, such that the magic vanishes there. This is not the only way in which magic can vanish, however, such that comparing magic and entanglement in different physical systems can help develop intuition about how these mysterious quantum properties are mutually complementary. To this end, we also display in figs.~\ref{fig:magic_plots_1} and~\ref{fig:magic_plots_2} the {\it concurrence} ${\cal \chi}$, that we already introduced in eq.~(\ref{chidef}). This varies between zero (no entanglement) and one (maximal entanglement) and is given explicitly for two-particle scattering by (see e.g. ref.~\cite{Gargalionis:2025iqs})
\begin{equation}
    \chi=2|a_{++}a_{--}-a_{+-}a_{-+}|,
    \label{Cdef}
\end{equation}
where $a_J$ is the coefficient of helicity state $\ket{J}$ in the normalised final state. We see that high entanglement indeed corresponds to low non-local magic. Furthermore, regions of maximal non-local magic correspond to intermediate entanglement. This is in keeping with the message of the Gottesmann--Knill theorem: maximally entangled states tend to be stabiliser, and thus can be just as efficiently simulated on a classical computer. However, {\it some} entanglement is needed in order to realise computational advantage, meaning that the amount of entanglement itself is not necessarily the best measure for how much ``quantumness'' of a quantum state is practically useful. This is captured instead by the non-local magic, and we note further that for 2-qubit systems the quantities can in fact be precisely related, as we have stated in eq.~(\ref{MABchi}). Strictly speaking, this means that the non-local magic is not in fact providing independent information from the concurrence. Nevertheless, the non-local magic remains crucial for two reasons. First, the fact that non-local magic and entanglement are precisely related will not necessarily be true for higher numbers of qubits. Secondly, the results of figs.~\ref{fig:magic_plots_1} and~\ref{fig:magic_plots_2} clearly show that it is not always easy to predict, from the concurrence plots alone, where the maximal non-local magic occurs as a function of the scattering angle. It is the latter that provides a genuine measure of the ``quantum usefulness'' of a particular state, making this measure a more useful way of conveying this information than trying to infer it from the concurrence. An analytic criterion describing when the magic is extremised in terms of the concurrence can be obtained using eq.~(\ref{MABchi}). Writing $\chi\equiv \chi(\theta)$, one has
\begin{equation}
    \frac{\partial {\cal M}_{AB}(\theta)}
    {\partial \theta}=\frac{2\chi(\theta)(1-2\chi^2(\theta))\chi'(\theta)}{\log(2)(\chi^4(\theta)-\chi^2(\theta)+1)}.
    \label{Mdiff}
\end{equation}
One then finds the following as a function of scattering angle:
\begin{enumerate}
    \item[(i)] The non-local magic is extremised for zero entanglement ($\chi=0$). This is a minimum, given that the non-local magic vanishes from eq.~(\ref{MABchi}).
    \item[(ii)] The non-local magic is extremised if the concurrence is too, for a given scattering angle. In particular, the non-local magic vanishes for maximum entanglement.
    \item[(iii)] The non-local magic is maximised if $\chi=1/\sqrt{2}$, which follows from eq.~(\ref{Mdiff}) and the related result
    \begin{displaymath}
    \left.\frac{\partial^2 {\cal M}_{AB}}{\partial\theta^2}
    \right|_{\chi=1/\sqrt{2}}
    =-\frac{16}{3}\frac{{\chi'}^2(\theta)}{\log(2)}<0.
    \end{displaymath}   
\end{enumerate}
The last point is particularly interesting given the implications of the Gottesmann-Knill theorem. If both zero and full entanglement do not convey any quantum computational advantage, there must be some non-maximal value of entanglement (as measured by the concurrence) that is optimal. This, then, is $\chi=1/\sqrt{2}$, a fact which was also noted in ref.~\cite{Liu:2025frx}.

\begin{figure}
    \begin{center}
        \includegraphics[width=\linewidth]{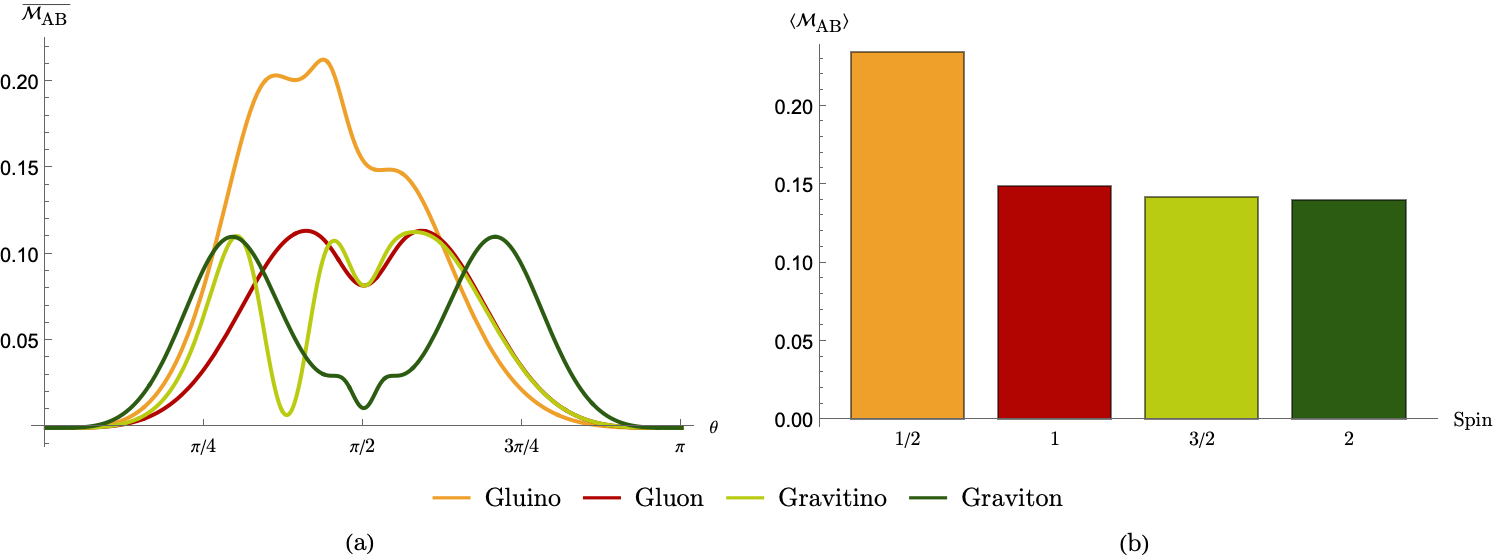}
        \caption{(a) The non-local magic power $\overline{\mathcal{M}_{AB}}$ of eq.~(\ref{magpownonlocal}) as a function of scattering angle $\theta$ and (b) the integrated non-local magic power $\left<\mathcal{M}_{AB}\right>$ of eq.~(\ref{intpow}) for gluinos, gluons, gravitinos, and gravitons.}
        \label{fig:magpownonlocal}
    \end{center}
\end{figure}

It is convenient to have a measure of the typical amount of magic associated with particles of different spin. One way to do this, as considered in ref.~\cite{Gargalionis:2025iqs}, is to calculate the {\it magic power}, namely the average amount of final state magic generated from all initial stabiliser states $\mathcal{S}$. We can do something similar for the non-local magic, such that we define the non-local magic power as
\begin{equation}
    \overline{{\cal M}_{AB}}=\frac{1}{60}\sum_{\xi\in{\cal S}}
    {\cal M}_{AB}(\ket{\psi(\xi)}),
    \label{MABpower}
\end{equation}
where $\ket{\psi(\xi)}$ denotes the final state associated with initial state $\xi$. Calculating this for each particle species yields the explicit results:
\begin{align}
\overline{{\cal M}_{AB}}\Big|_{\rm gluino}&=
-\frac{1}{30}\log_2\!\left(\frac{t^8 + 14 t^4 u^4 + u^8}{(t^2 + u^2)^4}\right)
-\frac{2}{15}\log_2\!\left(
\frac{16 t^8 + 32 t^6 u^2 + 20 t^4 u^4 + 4 t^2 u^6 + u^8}
{(2 t^2 + u^2)^4}
\right)\notag\\
&\quad-\frac{1}{15}\log_2\!\left(
\frac{
(4 t^4 - 12 t^3 u + 14 t^2 u^2 - 6 t u^3 + u^4)
(4 t^4 - 4 t^3 u + 2 t^2 u^2 - 2 t u^3 + u^4)
}{
(2 t^2 - 2 t u + u^2)^4
}
\right)\notag\\
&\quad-\frac{1}{15}\log_2\!\left(
\frac{
(4 t^4 + 4 t^3 u + 2 t^2 u^2 + 2 t u^3 + u^4)
(4 t^4 + 12 t^3 u + 14 t^2 u^2 + 6 t u^3 + u^4)
}{
(2 t^2 + 2 t u + u^2)^4
}
\right)\notag\\
&\quad-\frac{4}{15}\log_2\!\left(
\frac{16 t^8 + 16 t^6 u^2 + 24 t^4 u^4 + 4 t^2 u^6 + u^8}
{(2 t^2 + u^2)^4}
\right);\notag\\
\overline{{\cal M}_{AB}}\Big|_{\rm gluon}&=-\frac{4}{15}\log_2\!\left(
\frac{
4 s^4 u^4 (s^4 + t^4)^2
+ (s^4 + t^4)^4
+ 4 s^4 u^{12}
+ 2 u^8 (3 s^8 + 2 s^4 t^4 + 7 t^8)
+ u^{16}
}{
(s^4 + t^4 + u^4)^4
}
\right) \notag\\
&\quad-\frac{2}{15}\log_2\!\left(
\frac{
s^{16} + 14 s^8 (t^4 + u^4)^2 + (t^4 + u^4)^4
}{
(s^4 + t^4 + u^4)^4
}
\right)-\frac{1}{30}\log_2\!\left(
\frac{
t^{16} + 14 t^8 u^8 + u^{16}
}{
(t^4 + u^4)^4
}
\right) \notag\\
&\quad-\frac{1}{15}\log_2\!\left(
\frac{
s^{16} + 14 s^8 (t^2 - u^2)^4 + (t^2 - u^2)^8
}{
\bigl(s^4 + (t^2 - u^2)^2\bigr)^4
}
\right)\notag\\
&\quad-\frac{1}{15}\log_2\!\left(
\frac{
s^{16} + 14 s^8 (t^2 + u^2)^4 + (t^2 + u^2)^8
}{
\bigl(s^4 + (t^2 + u^2)^2\bigr)^4
}
\right); \notag\\
\overline{{\cal M}_{AB}}\Big|_{\rm gravitino}&=
-\frac{4}{15}\log_2\!\left(
\frac{
4 s^4 t^2 u^{18}
+ t^8 (s^4 + t^4)^4
+ 4 s^4 t^6 u^6 (s^4 + t^4)^2
+ 2 t^4 u^{12} (3 s^8 + 2 s^4 t^4 + 7 t^8)
+ u^{24}
}{
(s^4 t^2 + t^6 + u^6)^4
}
\right) \notag\\
&\quad -\frac{2}{15}\log_2\!\left(
\frac{
s^{16} t^8
+ 14 s^8 t^4 (t^6 + u^6)^2
+ (t^6 + u^6)^4
}{
(s^4 t^2 + t^6 + u^6)^4
}
\right)\notag \\
&\quad -\frac{1}{15}\log_2\!\left(
\frac{
s^{16} t^8
+ 14 s^8 t^4 (t^3 + u^3)^4
+ (t^3 + u^3)^8
}{
\bigl(s^4 t^2 + (t^3 + u^3)^2\bigr)^4
}
\right) \notag\\
&\quad -\frac{1}{15}\log_2\!\left(
\frac{
s^{16} t^8
+ 14 s^8 (t^4 - t u^3)^4
+ (t^3 - u^3)^8
}{
\bigl(s^4 t^2 + (t^3 - u^3)^2\bigr)^4
}
\right)\notag \\
&\quad -\frac{1}{30}\log_2\!\left(
\frac{
t^{24} + 14 t^{12} u^{12} + u^{24}
}{
(t^6 + u^6)^4
}
\right);\notag\\
\overline{{\cal M}_{AB}}\Big|_{\rm graviton}&=
-\frac{4}{15}\log_2\!\left(
\frac{
4 s^8 u^8 (s^8 + t^8)^2
+ (s^8 + t^8)^4
+ 4 s^8 u^{24}
+ 2 u^{16} (3 s^{16} + 2 s^8 t^8 + 7 t^{16})
+ u^{32}
}{
(s^8 + t^8 + u^8)^4
}
\right) \notag\\
&\quad -\frac{2}{15}\log_2\!\left(
\frac{
s^{32} + 14 s^{16} (t^8 + u^8)^2 + (t^8 + u^8)^4
}{
(s^8 + t^8 + u^8)^4
}
\right) \notag\\
&\quad -\frac{1}{15}\log_2\!\left(
\frac{
s^{32} + 14 s^{16} (t^4 - u^4)^4 + (t^4 - u^4)^8
}{
\bigl(s^8 + (t^4 - u^4)^2\bigr)^4
}
\right) \notag\\
&\quad -\frac{1}{15}\log_2\!\left(
\frac{
s^{32} + 14 s^{16} (t^4 + u^4)^4 + (t^4 + u^4)^8
}{
\bigl(s^8 + (t^4 + u^4)^2\bigr)^4
}
\right) \notag\\
&\quad -\frac{1}{30}\log_2\!\left(
\frac{
t^{32} + 14 t^{16} u^{16} + u^{32}
}{
(t^8 + u^8)^4
}
\right).
\label{magpownonlocal}
\end{align}
Upon substituting the results of eq.~(\ref{mandies2}), one may examine the non-local magic power as a function of scattering angle. This is shown in fig.~\ref{fig:magpownonlocal}, and one sees a somewhat complex dependence, that is smaller for particles of spin 1 and above, compared with gluinos (spin 1/2). One may condense the results into a single number for each spin by examining the integrated non-local magic power
\begin{equation}
    \langle {\cal M}_{AB}\rangle=\int_{0}^{\pi}
    d\theta {\cal M}_{AB}.
    \label{intpow}
\end{equation}
Results for each spin are also shown in fig.~\ref{fig:magpownonlocal}, and confirm the fact that the typical amount of non-local magic is much smaller for spins of 1 and greater. As is the case for local magic in the helicity basis, the integrated non-local magic power decreases monotonically with increasing spin ~\cite{Gargalionis:2025iqs}, a conclusion which may be portable to other quantum systems.

\section{Deforming Yang-Mills theory}
\label{sec:EFT}

In the previous section, we have seen that there are good physical motivations for using the helicity basis, in that for many final states, including those corresponding to polarised collider initial states, the non-local magic is manifest. In order to investigate how special this situation is, we now focus purely on qubits of spin one (gluons), but deform their underlying Yang-Mills theory. To this end, one may replace the Yang-Mills Lagrangian ${\cal L}_{\rm YM}$ as follows:
\begin{equation}
    {\cal L}_{\rm YM}\rightarrow {\cal L}_{\rm YM}+{\cal L}_{F^3},
    \quad {\cal L}_{F^3}=\frac{c}{\Lambda^2}f_{abc} {F^{a\mu}}_\nu {F^{b\nu}}_{\rho}{F^{c\rho}}_\mu.
    \label{LF3}
\end{equation}
The additional operator naturally arises in new physics extensions to the Standard Model of particle physics, given that all such extensions can be expanded as an effective field theory if the energy scale of the new physics ($\Lambda$) is sufficiently high compared to collider energies. The operator is built from three field strength tensors, and any given new physics theory would fix the so-called {\it Wilson coefficient} $c$. Here, our motivation for considering this operator is to probe whether or not our previous conclusions about the helicity basis are robust against deformations of the theory. 

Tree-level (leading order) gluon amplitudes including the operator of eq.~(\ref{LF3}) were first obtained in ref.~\cite{Dixon:1993xd}, and are collected here for convenience in appendix~\ref{app:amplitudes}. They do not change the existing amplitudes appearing in eq.~(\ref{ampmatrix}), but instead lead to new non-zero off-diagonal components. In order to probe whether the helicity basis continues to manifest the non-local magic for realistic (polarised) collider initial states, it is sufficient to consider the initial state $\ket{+-}$, which is in group IV($\otimes$) in table~\ref{tab:stabstates} (n.b. results for $\ket{-+}$ will be similar; initial states $\ket{++}$, $\ket{--}$ in group II($\otimes$) do not generate final state non-local magic for pure gluon theory). 
We then require the additional amplitudes
\begin{displaymath}
{\cal A}(+-\rightarrow ++),\quad {\cal A}(+-\rightarrow --),
\end{displaymath}
and an explicit calculation of the concurrence in the final state (from eq.~(\ref{Cdef})) yields
\begin{equation}
    \chi_{{\rm YM}+F^3}(\ket{+-})=
    \frac{t^2u^2|4g^4-\tilde{c}^2 s^2|}
    {\tilde{c}^2s^2t^2u^2+2g^4(t^4+u^4)},
    \label{CF3}
\end{equation}
where for brevity we have defined the dimensionful combination
\begin{equation}
\tilde{c}=\frac{3c}{\Lambda^2}.
\label{ctildedef}
\end{equation}
From the concurrence, one may find the non-local magic according to eq.~(\ref{MABchi}):
\begin{equation}
  {\cal M}_{AB}^{{\rm YM}+F^3} = -
\log_2\!\left(
1 - 
\frac{
4 g^{4} t^{4} u^{4} (t^{2}+u^{2})^{2} (\tilde{c}^{2} s^{2}-4g^{4})^{2}
\left( \tilde{c}^{2} s^{2} t^{2} u^{2} + g^{4} (t^{2}-u^{2})^{2}\right)
}{
\left( \tilde{c}^{2} s^{2} t^{2} u^{2} + 2g^{4} (t^{4}+u^{4})\right)^{4}
}
\right).
\label{MABF3}
\end{equation}
Furthermore, the local magic in the final state is found to be
\begin{equation}
 {\cal M}_2^{{\rm YM}+F^3} = -
\log_2\!\left(
\frac{
\, \tilde{c}^{8} s^{8} t^{8} u^{8}
+ 28\, \tilde{c}^{4} g^{8} s^{4} t^{4} u^{4} \left(t^{8} + 6 t^{4} u^{4} + u^{8}\right)
+ 16g^{16} \left(t^{16} + 14 t^{8} u^{8} + u^{16}\right)
}{
\left( \tilde{c}^{2} s^{2} t^{2} u^{2} + 2g^{4} \left(t^{4} + u^{4}\right)\right)^{4}
}
\right).
\label{M2F3}
\end{equation}
One may then surmise that, whilst eqs.~(\ref{MABF3}, \ref{M2F3}) agree in the pure YM limit of $\tilde{c}\rightarrow 0$, they do not agree for non-zero values of the Wilson coefficient (even after eliminating one of the Mandelstam invariants using eq.~(\ref{momcon})). Thus, deforming Yang-Mills theory with the operator of eq.~(\ref{LF3}) breaks the property that the helicity basis makes the non-local magic manifest for polarised collider initial states. 

\begin{figure}
\begin{center}
    {\includegraphics[width=\linewidth]{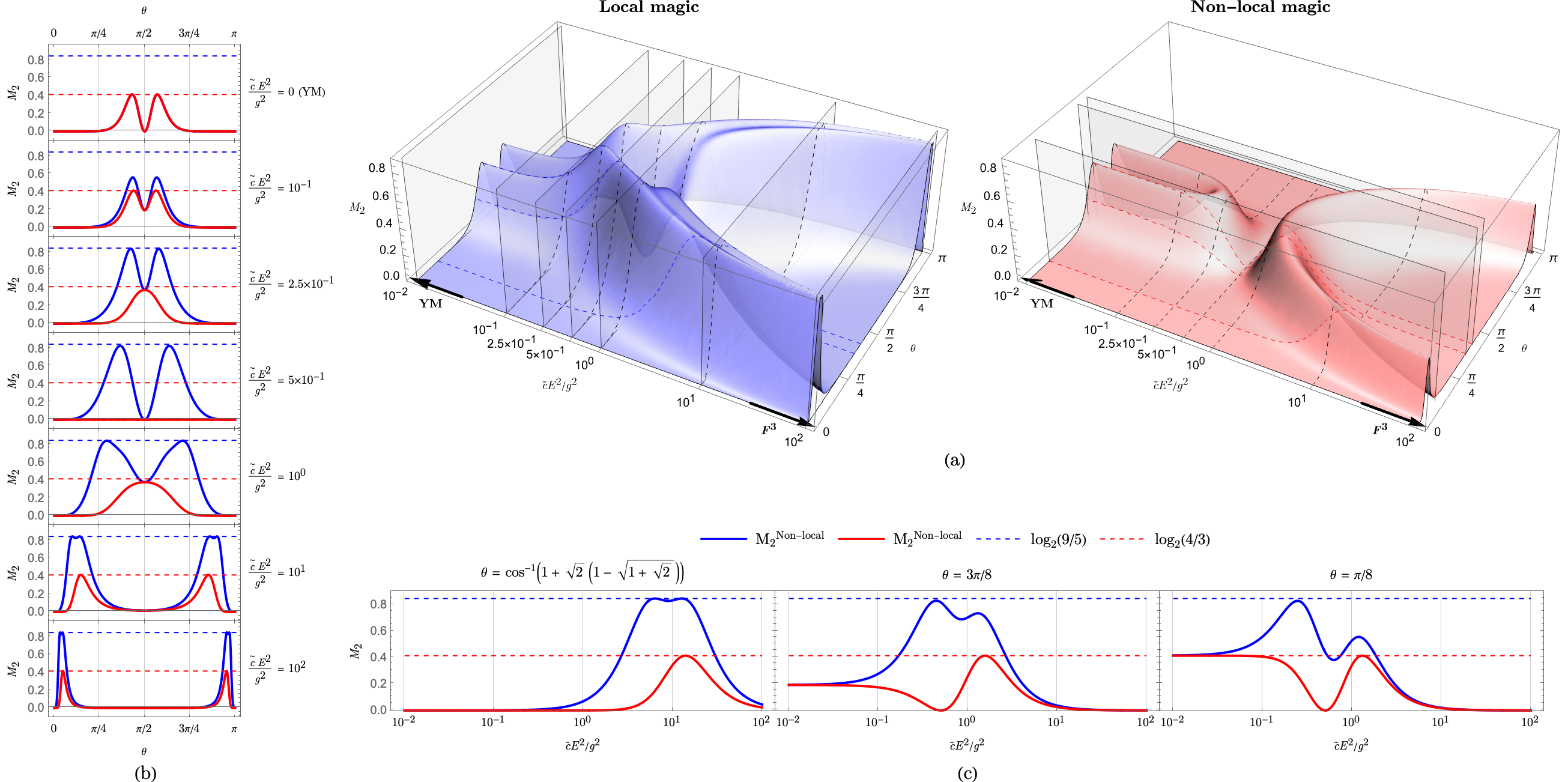}}
    \caption{Final state local (blue) and non-local (red) magic in the helicity basis corresponding to the initial state $\ket{+-}$. Results are shown for the deformed Yang-Mills theory of eq.~(\ref{LF3}) in (a) as a function of the scaled Wilson coefficient $\tilde{c}$ of eq.~(\ref{ctildedef}) and scattering angle $\theta$. Cross-sections are shown for particular values of $\tilde c$ in (b) and $\theta$ in (c), with the maximum analytic magic values for this scattering process. The section with $\theta=\cos^{-1}\left[
    1+\sqrt 2\left(1-\sqrt{1+\sqrt2}\right)\right]$ corresponds to the maximum non-local magic in the Yang-Mills limit.}
    \label{fig:F3plots}
\end{center}
\end{figure}

We can see this more clearly in fig.~\ref{fig:F3plots}, which displays results for quantities of eqs.~(\ref{MABF3}, \ref{M2F3}) as functions of the scaled Wilson coefficient of eq.~(\ref{ctildedef}). By comparing two surfaces, one can clearly see that results for (helicity basis) local, and non-local, magic are different in general. Furthermore, the presence of the operator in eq.~(\ref{LF3}) substantially enhances the local magic, where this effect becomes more pronounced as the Wilson coefficient increases. The maximum magic we find can be derived upon noting that the final state obtained from $\ket{+-}$ is of the form
\begin{equation}
    \ket{\psi}=a\ket{++}+b\ket{+-}+c\ket{-+}+a\ket{--}.
    \label{psiform2}
\end{equation}
From eq.~(\ref{Mqdef}), the SSRE is given by
\begin{equation}
M_2(\ket{\psi})=2-\log_2\left(\sum_P \langle \psi|P|\psi\rangle^4\right),
    \label{M2val}
\end{equation}
such that the magic will be maximised when the sum over the fourth powers of Pauli expectation values is minimised, subject to the constraint that $\ket{\psi}$ be normalised, which amounts to
\begin{equation}
    2a^2+b^2+c^2=1.
    \label{abcnorm}
\end{equation}
The sum over Pauli strings in eq.~(\ref{M2val}) evaluates to
\begin{equation}
\sum_P\langle \psi|P|\psi\rangle^4= 4(16a^8 + 28a^4b^4 + b^8 + 168a^4b^2c^2 + 28a^4c^4 + 14b^4c^4 + c^8),
\label{paulisum}
\end{equation}
and minimising this with the above constraint yields
\begin{equation}
\left(\sum_P\langle \psi|P|\psi\rangle^4\right)_{\rm min.}
=\frac{20}{9},
 \label{psummin}
\end{equation}
with example parameter values
\begin{equation}
 a_*=-\frac{1}{2\sqrt{3}},\quad b_*=-\frac{\sqrt{3}}{2},\quad
 c_*=\frac{1}{2\sqrt{3}}.
 \label{abcmin}
\end{equation}
The maximum magic is then
\begin{equation}
M_2^{\rm max.}=\log_2\left(\frac95\right)\simeq 0.848,
    \label{M2max}    
\end{equation}
which we display in fig.~\ref{fig:F3plots}. It can clearly be seen that this maximum value is indeed saturated. It remains somewhat less, however, than the overall possible maximum for local magic proposed in ref.~\cite{Liu:2025frx}, which in our conventions is $\log_2(16/7)\simeq 1.19$\footnote{Reference~\cite{Liu:2025frx} defines magic using $\log_e$ rather than $\log_2$, such that our numerical value will be different.}. Similarly, we show the maximum possible value of non-local magic, which can be straightforwardly obtained from eq.~(\ref{MABexact}) as $\log_2(4/3)\simeq 0.415$. In this case, the ultimate theoretical upper bound is indeed saturated by gluon scattering in the (deformed) YM theory.

In figure~\ref{fig:F3plots}(c), we show the (non-)local magic as a function of the scaled Wilson coefficient for a set of fixed scattering angles. We again see that the effect of the deformation is to create a sizeable difference between the local and non-local magic. This stresses the importance of using the non-local magic to quantify non-stabiliserness if applying quantum information measures as probes of new physics (see e.g. refs.~\cite{Fabbrichesi:2025ywl,Aoude:2025jzc}): the effect of the new physics may be such that the local magic in the helicity basis no longer matches the non-local magic.

\section{Conclusion}
\label{sec:conclude}

The study of quantum information measures in different physical systems continues to attract attention, due to applications to quantum computing, as well as the possibility to generate insights using one physical scenario that may prove relevant for others. A particular measure of interest is {\it non-stabiliserness} or {\it magic}, which measures the ability of a quantum state to provide genuinely useful quantum advantage as part of a computational algorithm. Magic is also a crucial ingredient in designing fault-tolerant quantum computers, such that how to produce and enhance magic in different systems has become a wide-ranging topic of study.

In this paper, we have studied magic production in the $2\rightarrow 2$ scattering of massless particles of different spin. Typically,  measures of the magic of the final quantum state depend on the basis used to map spin directions of the particles to physical space. A common choice is the helicity basis, and the magic for the scattering processes considered here was already examined in this basis in ref.~\cite{Gargalionis:2025iqs}, quantified using the Second Stabiliser R\'{e}nyi Entropy (SSRE) of ref.~\cite{Leone:2021rzd}. Recently, basis-independent measures of magic have appeared, such as the non-local non-stabiliserness (or non-local magic) of ref.~\cite{Qian:2025oit}. This involves minimising the local magic over all possible local unitary transformations of each individual qubits, and there is then no guarantee that any one basis is such that the non-local magic is manifest. We have performed a detailed study of the non-local magic for the scattering of gluinos (spin 1/2), gluons (spin 1), gravitinos (spin 3/2) and gravitons (spin 2). We have derived analytic results for the non-local magic in each case, including intermediate results that do not depend on the spin of the particle in question. We have grouped initial stabiliser states into different (sub-)groups, according to the profile of (non-)local magic in the corresponding final states. These groupings are the same for all spins, where differences between the profiles are solely explained in terms of the additional angular dependence introduced either by supersymmetric Ward identities or the KLT relations. 

Based on this analysis, we find that the helicity basis is such that 18 out of the 60 possible initial stabiliser states have local magic in the final state that is the same as the non-local magic. Perhaps more importantly, initial states that would actually arise in polarised collider experiments (i.e. product states in the computational basis) have manifest non-local magic in the helicity basis. Our results further allow us to examine the typical amount of non-local magic produced as a function of the spin of the (massless) scattering particles, which we have quantified using the non-local magic power (average non-local magic produced from all stabiliser initial states). As in ref.~\cite{Gargalionis:2025iqs}, the integrated non-local magic power is a monotonically decreasing function of the spin, and falls sharply upon considering spins greater than 1/2. This conclusion may perhaps be portable to condensed matter systems, or indeed other physical systems of relevance for quantum computing.

The above conclusions do not generalise if one moves beyond standard Yang-Mills theory, and its closely related cousins (e.g. pure gravity and supersymmetric extensions). To examine this, we deformed Yang-Mills theory by adding a higher dimensional operator, as would typically arise in generic new physics scenarios. Interestingly, the effect of this operator is to enhance the local magic in the helicity basis, in such a way that a significant difference with the non-local magic occurs. This underlines the importance of considering basis-independent measures of non-stabiliserness if using such measures as probes of new physics.  

The study of non-stabiliserness in high energy physics remains a relatively recent development, and we hope that our analysis contributes further useful insights to this fascinating interdisciplinary endeavour.

\section*{Acknowledgments}

CDW, NM and MRA are supported by the UK Science and Technology Facilities Council
(STFC), including through the Consolidated Grant ST/P000754/1 ``String theory, gauge theory
and duality''. MJW is supported by the Australian Research Council
Discovery Project DP220100007, and MJW and JG are both supported by the ARC Centre of Excellence for Dark Matter Particle Physics (CE20010000). ENVW acknowledges the support he has received for this research through the provision of an Australian Government Research Training Program Scholarship.

\appendix

\section{Helicity amplitude combinations for different particle species}
\label{app:amplitudes}

In eq.~(\ref{ampmatrix2}), we write the structure of the amplitude matrix for $2\rightarrow 2$ scattering in terms of coefficients obtained from the relevant helicity amplitudes, and which make the symmetry of the matrix manifest. For completeness, we here collect the results for different particle species. For gluinos and gluons, one has
\begin{align}
\Gamma^{(1/2)}=-2g^2\left(\frac{F_2 s}{u}+\frac{F_1 t}{u}\right),
\quad
\Sigma^{(1/2)}=-2g^2\left(F_1+
\frac{F_2 s}{t}\right),\quad
\Delta^{(1/2)}=-2g^2\left(\frac{F_2 s}{u}+\frac{F_1 t}{u}\right)
\label{amps1/2}
\end{align}
and
\begin{align}
\Gamma^{(1)}=-2g^2\left(\frac{F_1 s}{t}+\frac{F_2 s}{u}\right),
\quad
\Sigma^{(1)}=-2g^2\left(\frac{F_1 u^2}{st}+
\frac{F_2 u}{s}\right),\quad
\Delta^{(1)}=-2g^2\left(\frac{F_1 t}{s}+\frac{F_2 t^2}{su}\right)
\label{amps1}
\end{align}
respectively, where $g$ is the coupling constant, and we have labelled coefficients according to their spin. Furthermore, $F_1$ and $F_2$ are independent combinations of colour structure constants, which turn out to cancel when calculating the (non-)local magic. Corresponding results for gravitinos and gravitons are
\begin{align}
\Gamma^{(3/2)}=-\left(\frac{\kappa}{2}\right)^2
\frac{s^2 t}{u^2},
\quad
\Sigma^{(3/2)}=-\left(\frac{\kappa}{2}\right)^2 u,\quad
\Delta^{(3/2)}=-\left(\frac{\kappa}{2}\right)^2\frac{t^3}{u^2}
\label{amps3/2}
\end{align}
and
\begin{align}
\Gamma^{(2)}=-\left(\frac{\kappa}{2}\right)^2
\frac{s^3 }{tu},
\quad
\Sigma^{(2)}=-\left(\frac{\kappa}{2}\right)^2 \frac{u^3}{st},\quad
\Delta^{(2)}=-\left(\frac{\kappa}{2}\right)^2\frac{t^3}{su},
\label{amps2}
\end{align}
where $\kappa=\sqrt{32\pi G_N}$ is the gravitational coupling constant in terms of Newton's constant $G_N$.

In section~\ref{sec:EFT}, we also need extra contributions to gluonic amplitudes arising from the presence of the additional interaction of eq.~(\ref{LF3}). The relevant results were first obtained in ref.~\cite{Dixon:1993xd}. They do not change the existing amplitudes in eq.~(\ref{ampmatrix2}), but instead lead to extra non-zero values of the amplitudes matrix. The specific helicity amplitudes we need  (with incoming initial momenta) are
\begin{equation}
    {\cal A}(1^+,2^-,3^+,4^+)=
    {\cal A}(1^+,2^-,3^-,4^-)=
    -\frac{3c}{\Lambda^2}(F_1 u+F_2 t).
\end{equation}

% =================== Bibliography ===================
\bibliographystyle{utphys}
\bibliography{refs}

\end{document}